\documentclass[%
reprint,
twocolumn,
superscriptaddress,
preprintnumbers,
nobibnotes,
nofootinbib,
 amsmath,amssymb, 
 aps, prl,
 longbibliography,
]{revtex4-1}

\usepackage{cancel}
\usepackage{accents}
\usepackage{mciteplus,slashed}
\usepackage{amssymb,cancel,amsmath}
\usepackage{dcolumn}
\usepackage{bm}
\usepackage{soul}
\usepackage[caption=false]{subfig}
\usepackage{appendix}
\usepackage{physics}
\usepackage{booktabs}
\usepackage{comment}
\usepackage[T1]{fontenc}	
\usepackage{csvsimple}
\usepackage[bookmarksnumbered=true]{hyperref} 
\hypersetup{
    colorlinks = true,
    linkcolor = blue,
    anchorcolor = blue,
    citecolor = blue,
    filecolor = blue,
    urlcolor = blue
    }
\usepackage[section]{placeins}
\usepackage[capitalise]{cleveref}
\usepackage{booktabs}
\usepackage{graphicx}
\usepackage{mathrsfs}
\graphicspath{{Figures/}}
\AtBeginDocument{
\heavyrulewidth=.08em
\lightrulewidth=.05em
\cmidrulewidth=.03em
\belowrulesep=.65ex
\belowbottomsep=0pt
\aboverulesep=.4ex
\abovetopsep=0pt
\cmidrulesep=\doublerulesep
\cmidrulekern=.5em
\defaultaddspace=.5em
}
\usepackage[dvipsnames]{xcolor}
\usepackage[normalem]{ulem}
\usepackage{fontawesome} 
\usepackage[force]{feynmp-auto}
\usepackage{bbm}
\def\iu{\mathrm{i}}
\def\e{\mathrm{e}}

\def\nMeas{\aleph}

\definecolor{interorange}{RGB}{1.0,0.3098,0}
\definecolor{forestgreen}{HTML}{228B22}

\begin{document}

\title{Coherent collisional decoherence }

\author{Leonardo Badurina}
\email{badurina@caltech.edu}
\affiliation{Walter Burke Institute for Theoretical Physics, California Institute of Technology, Pasadena, CA 91125, USA}
\author{Clara Murgui}
\email{cmurgui@ifae.es}
\affiliation{Departament de F\'isica, Universitat Autonoma de Barcelona, 08193 Bellaterra, Barcelona, Spain}
\author{Ryan Plestid}
\email{rplestid@caltech.edu}
\affiliation{Walter Burke Institute for Theoretical Physics, California Institute of Technology, Pasadena, CA 91125, USA}

\date{\today}

\preprint{CALT-TH/2024-005}

\begin{abstract}
    We study the decoherence of a system of $N$ non-interacting heavy particles (atoms) due to coherent scattering with a background gas. We introduce a framework for computing the induced phase shift and loss of contrast for arbitrary preparations of $N$-particle quantum states. We find phase shifts that are inherently $(N\geq 2)$-body effects and may be searched for in future experiments. We analyze simple setups, including a two-mode approximation of an interferometer. We study fully entangled $N00N$ states, which resemble the correlated positions in a matter interferometer, as well as totally uncorrelated product states that are representative of a typical state in an atom interferometer.  
    We find that the extent to which coherent enhancements increase the rate of decoherence depends on the observable of interest, state preparation, and details of the experimental design. 
    In the context of future ultralow-recoil (e.g., light dark matter) searches with atom interferometers we conclude that: {\it i}) there exists a coherently enhanced scattering phase which can be searched for using standard (i.e., contrast/visibility and phase) interferometer observables;  {\it ii}) although decoherence rates of one-body observables are {\it not} coherently enhanced, a coherently enhanced loss of contrast can still arise from dephasing; and {\it iii}) higher statistical moments (which are immediately accessible in a counting experiment) {\it are} coherently enhanced and may offer a new tool with which to probe the soft scattering of otherwise undetectable particles in the laboratory. 
\end{abstract}

 \maketitle

\section{Introduction}
Decoherence is a pervasive quantum phenomenon which underlies much of what we understand about the quantum to classical transition \cite{Zurek:2003zz,Hornberger:2006xxx}. 
It is universal, and can occur in kinematic regimes in which no other measurable effect would be induced.
For example, {\it collisional decoherence} arises from the ultra-soft scattering of a probe with a quantum system (e.g., $N$ non-interacting\footnote{By ``non-interacting atoms'' we mean that interactions between atoms can be neglected in comparison with the interactions between the atoms and the environment.} atoms), which leaves the positions of single atoms essentially unchanged as a result of a momentum transfer $\vb{q}$ \cite{Gallis:1990,Hornberger:2003}. The  position is resolved within a region of size $1/{|\vb{q}|}$. If the resolution size is smaller than the separation of the spatial superposition, i.e., $1/{|\vb{q}|} < |\mathbf{\Delta x}|$, then its quantum state decoheres, and position is einselected as the preferred classical label \cite{Zurek:1981xq,Gallis:1990,Hornberger:2003}.
This phenomenon is generic, it can be induced by almost any scattering process and its observable signatures are calculable, being directly related to the 
scattering cross section \cite{Joos:1984uk,Diosi:1994xj,Hackermuller:2004xxx,Hornberger:2006,Adler:2006gt}.

Collisional decoherence is therefore intimately tied to the limit of soft scattering. This has lead Riedel to propose, in a series of papers~\cite{Riedel:2012ur,Riedel:2015pxa,Riedel:2016acj}, that measurements of decoherence (e.g., with matter or atom interferometers)  can be used as a sensitive probe of soft spin-independent scattering induced by sub-GeV dark matter. This idea has been further pursued in more recent literature~\cite{Du:2022ceh,Du:2023eae,Kaltenbaek:2023xtz}.
Traditional direct detection is subject to detector thresholds, and even large scattering cross sections can be unobservable if their reaction products are invisible (i.e., lie below detection thresholds); this makes models dominated by soft-scattering particularly challenging to search for. Atom interferometers are effectively threshold-less detectors, and therefore offer a complimentary probe of models with large cross sections, but whose scattering against ordinary matter is dominated by small momentum transfers. These momentum transfers can be so small that atom kinematics are negligibly affected, and  effects such as atom loss\footnote{Atom loss includes both getting ``kicked out'' of the cloud, or being Doppler-shifted outside the velocity class of an experimentally relevant laser transition.} can be completely neglected. A sample of models which naturally satisfy this criteria are discussed in Ref.~\cite{Riedel:2016acj}.

In the limit of low momentum transfer, it is natural to consider enhanced sensitivity that would arise due to the constructive interference between scattered waves from different atoms. In the context of scattering from nuclei \cite{Freedman:1973yd,Tsai:1973py,Engel:1992bf} this phenomenon is often referred to as {\it coherent scattering} and refers specifically to scattering rates that scale as $N^2$ as opposed to $N$ (as would be obtained from an incoherent sum of scattering cross sections from individual atoms). In the rest of this paper we will use the term ``coherent scattering'' or ``coherent enhancement'' in this specific technical sense of parametric scaling, i.e., $N^2$ vs $N$.
Coherent enhancements are commonplace in scattering off nuclei \cite{Bednyakov:2018mjd}, and offer order of magnitude improvements in sensitivity for direct detection experiments when considering models of spin-independent dark matter \cite{Engel:1992bf,Feng:2010gw}.
In the context of matter and atom interferometers, coherent enhancements are quite dramatic since $N\sim 10^6-10^{10}$ in some experiments~\cite{Frye:2021,Abe:2021, Overstreet:2022,Kaltenbaek:2015kha}. 
It is therefore crucial to reliably establish how coherent scattering enters into the derivation of, and potentially modifies, formulas for collisional decoherence. 


Surprisingly no such formalism currently exists. 
The original papers on collisional decoherence \cite{Joos:1984uk,Gallis:1990} and subsequent publications~\cite{Hornberger:2003, Vacchini_2009} all explicitly worked on an atom-by-atom basis.
In this work, we address this lacuna by developing a general and flexible formalism for computing collisional decoherence for a non-interacting $N$-body system. As we will see, a consistent derivation of decoherence for an $N$-body system differs from $N$ iterations of a single-body experiment. 

Before proceeding to technical details, let us sketch a brief motivation as to why $N$-particle coherent scattering may differ from $N$ independent single-particle experiments. 
Consider an interferometer, with a left arm and a right arm. 
Suppose an initial state is prepared such that every atom will end up in a superposition of left and right arms at a later time, i.e.,\ ${\ket{\Psi} = \bigotimes_{i=1}^N \tfrac1{\sqrt2}(\ket{L_i}+\ket{R_i})}$.  
The wavefunction will have many branches with varying numbers of particles in the left and right arms of the interferometer, respectively $N_L$ and $N_R$.  
Coherent scattering with the background gas will modify the coefficients that multiply each branch of the wavefunction in a manner that depends on $N_L$ and $N_R$. 
Thus the scattered branch of the wavefunction will no longer be a product state and will be entangled with the environment in an $N_L$-dependent way.
When considering measurements of {\it only} the atoms, one may trace out the environment. Since the environment and atoms are entangled in an $N$-dependent fashion, one expects the rate of decoherence to depend on $N$.
The purpose of this paper is to provide a formalism for computing these effects.  In what follows we compute collisional decoherence for a non-interacting $N$-body system.
The formalism is general and flexible, and can easily interpolate between different limits.

Our results provide a conclusive answer to the question of when and how coherent enhancements arise in the context of atom and matter interferometers. 
We find that, at the level of the $N$-particle density matrix, coherent scattering {\it always} influences the rates of decoherence and induces a coherently enhanced phase shift. This is a necessary, but not sufficient, condition for coherent enhancements to be observable in the lab. In particular we find that when restricted to one-body measurements, only the coherently enhanced phase survives. Nevertheless, given an experiment with many iterations it is trivial to construct higher-body observables by considering statistical fluctuations (i.e., higher order moments). We find that these quantities generically are sensitive to enhanced rates of decoherence for generic state preparations. 
We stress that this observation is not restricted to ``exotic'' states that are delicately prepared in the laboratory (e.g. $N00N$ states), since it also applies to objects whose constituents are entangled by inter-particle interactions i.e, for matter interferometers. 

The rest of the paper is organized as follows: In \cref{sec:coll-deco}, we develop a formalism for collisional decoherence from an $N$-body system in the limit of small momentum transfers. We work in terms of a general $T$-matrix, and then specialize our analysis to weak coupling where the Born approximation applies. Next, in \cref{sec:Two-Arm}, we consider a toy-model of a two-arm interferometer. We first discuss product states in \cref{sec:Product-States} (for atom interferometers), focusing on when and how coherently enhanced scattering rates influence relevant observables. We discuss statistical fluctuations of observables in \cref{sec:Stat-Fluct}, which can be sensitive to coherent enhancements. We find that strong rates of collisional decoherence that respect a permutation symmetry can lead to non-classical statistics at late times.  In \cref{sec:Entangled}, we consider  entangled states (for matter interferometers).
Finally, in \cref{sec:Disc-Concl}, we summarize our findings,  contextualize our results, and comment on future directions. 

\section{Collisional decoherence \label{sec:coll-deco}} 
A ``probe'' with momentum $\vb{p}$, $\ket{\pi(\vb{p})}$, is incident and scatters upon a gas of $N$ atoms prepared in a state $\rho_A$. The total Hamiltonian can be written as $\hat{H}=\hat{H}_A+\hat{H}_\pi+ \hat{H}_{\rm int}$, where $\hat{H}_A$ acts only on the atoms' Hilbert space, $\hat{H}_\pi$ only on the probe's Hilbert space, and $\hat{H}_{\rm int}$ is the interaction between the two systems. As discussed in \cref{app:pot_scatt}, in the limit of $|\vb{q}| /M_A\rightarrow 0$, where $M_A$ is the mass of the atom, we can model $\hat{H}_{\rm int}$ as a sum of static potentials. For example, if $\pi$ interacts with atoms via a massive spin-1 mediator (analogous to a photon with non-vanishing mass) then $\hat{H}_{\rm int} \simeq \int \dd^3 y  \sum_i V(\vb{y}-\vb{x}_i) \hat{n}_\pi(\vb{y})$ where, $\{\vb{x}_i\}$ are the location of atoms, $\hat{n}_\pi(\vb{x})$ is the number density operator, and $V(\vb{x})$ is a Yukawa potential with a range set by the mass of the mediator. We assume that  $\hat{H}_A$ is a  non-relativistic Hamiltonian, while $\hat{H}_\pi$ may be either relativistic or non-relativistic. 

Prior to scattering, the system and probe are described in terms of the total density matrix $\rho= \rho_{A} \otimes \rho_\pi$; we will assume $\rho_\pi$ is a mixed state diagonalized in momentum space.
The dynamics of collisions between the atoms and probe particles are captured by the (unitary) scattering operator $S$. Assuming that the probe gas is sufficiently dilute, so that the time between collisions is long compared to the duration of a collision, we  
model the effects of collisions by a two-particle scattering operator 
$\rho\rightarrow \rho' = S \rho S^\dagger$. 

Using the standard definition of the $T$-matrix, $S=1 +\iu \, T$, we may write the change in the density matrix due to scattering, $\Delta \rho=\rho'-\rho$ as 
\begin{equation}
    \begin{split}
       \Delta \rho &=  \tfrac{\iu}{2} \qty[ T+T^\dagger,\rho] + \tfrac{\iu}{2} \qty{T- T^\dagger, \rho} +  T \rho T^\dagger \\
                &= \tfrac{\iu}{2} \qty[ T+T^\dagger,\rho] - \tfrac12\qty{ T^\dagger T, \rho} +  T \rho T^\dagger ~,
    \end{split}
\end{equation}
where in going to the second line we used the optical theorem, $\iu \, (T-T^\dagger) = -T^\dagger T$. The above equation exhibits a Lindbladian form. The anti-commutator piece describes the unitary dynamics and may cause phase-shifts in the evolution of the density matrix. The second and third terms are responsible for decoherence, where $T$ can be identified as the jump operators.
Our goal is to compute the reduced density matrix of the atoms 
\begin{equation}
    \begin{split}
        \rho_{A}' &=  {\rm Tr}_\pi \rho'~ = \int\frac{\dd^3p}{(2\pi)^3} \bra{\pi(\vb{p})}\rho'\ket{\pi(\vb{p})}~,
    \end{split}
\end{equation}
where $\mathbf{p}$ is the three-momentum of the probe and ${\rm Tr}_\pi$ is the trace over environmental degrees of freedom. 
In the static limit (defined by $|\vb{q}|\ll M_A$, where $M_A$ is the atomic mass), atomic position is conserved in the scattering process and $\ket{\{\vb{x}\}} = \bigotimes_{i=1}^N \ket{\vb{x}_i}$ acts as a good 
pointer basis \cite{Zurek:1981xq,Gallis:1990,Busse:2009xxx}. At the level of the $T$-matrix, this manifests as 
\begin{equation}
        T \left ( \ket{\{\vb{x}\}} \otimes \ket{\pi} \right )= \ket{\{\vb{x}\}}\otimes \qty(T_{\{\vb{x}\}} \ket{\pi})~.
\end{equation}
We refer to $T_{\{\vb{x}\}}$ as the induced $T$-matrix; it depends on all of the atomic spatial coordinates. Whenever the probe interacts in the same way with all of the atoms, $T_{\{\vb{x}\}}$ is invariant under permutations of atomic positions. It acts on the probe Hilbert space and has matrix elements  normalized as is appropriate for potential scattering, i.e.,~\cite{Berestetskii:1982qgu}
\begin{equation}\label{eq:T-matrix-M}
    \mel{\pi(\vb{p}')}{T_{\{\vb{x}\}}}{\pi(\vb{p})} = (2\pi)\delta(E'-E)\mathcal{M}_{\{\vb{x}\}}(\vb{p}',\vb{p}) \, ,
\end{equation}  
being $E$ and $E'$ the energies of the initial and final states, respectively. It may be computed by treating each atom as a background potential centered at $\vb{x}_i$ and summing the resulting amplitudes (\textit{cf.} \cref{app:scatt_conventions} for a discussion on scattering theory conventions and \cref{app:pot_scatt} for a derivation). 

We may expand the scattering matrix elements perturbatively using the Born series. For feebly interacting particles, such as dark matter or neutrinos, this treatment is always justified (even when considering matter interferometers). For particles with larger interaction rates, e.g., photons or neutrons, the Born approximation applies in the dilute limit where $na^3\ll1$ where $n$ is the number density and $a$ is the probe-atom scattering length \cite{abrikosov2012methods,Braaten:1996rq,Hammer:2000xg}.  
At first order we have 
\begin{equation}\label{eq:first-order-m-element}
    \mathcal{M}_{\{\vb{x}\}}^{(1)}(\vb{p}',\vb{p}) = \sum_i \widetilde{V} (\vb{q}) \e^{\iu \vb{q} \cdot \vb{x}_i}~,
\end{equation}
where $\widetilde{V}(\vb{q})$ is the Fourier transform of the scattering potential $V(\vb{x})$ with $\vb{q}=\vb{p}'-\vb{p}$. At second order we have 
\begin{equation}\label{eq:second-order-m-element}
    \begin{split}
    \mathcal{M}_{\{\vb{x}\}}^{(2)}(\vb{p}',\vb{p}) = & \sum_{ij}  \int \frac{\dd^3 q_1}{(2\pi)^3} \frac{\dd^3 q_2}{(2\pi)^3}\e^{\iu \vb{q}_1 \cdot \vb{x}_j}\e^{\iu \vb{q}_2 \cdot \vb{x}_i} \\
    &
    \quad \times \widetilde{V} (\vb{q}_2)  G_\pi(\vb{p}+\vb{q}_1)\widetilde{V} (\vb{q}_1) \\
    &
    \quad \hspace{0.1\linewidth}
    \times (2\pi)^3\delta^{(3)}(\vb{q}_1+\vb{q}_2 - \vb{q}) \, ,
    \end{split}
\end{equation}
where the probe propagator appears as $G_\pi$. 

Without loss of generality we can write any atomic density matrix in terms of its position eigenstates 
\begin{equation}\label{eq:rho_A}
    \rho_A = \int [\dd \{\vb{x}\}] [\dd \{\vb{x}'\}] ~\rho_A\qty(\{\vb{x}\},\{\vb{x}'\})~\ket{\{\vb{x}\}} \bra{\{\vb{x}'\}}~\, ,
\end{equation}
where $[\dd \{\vb{x}\}]=\prod_{i = 1}^{N} \dd^3x_i$.
The behavior of the states $\ket{\{\vb{x}\}}$ as a pointer basis allows us to evolve each of these matrix elements independently of one another. The matrix elements of  $\Delta \rho_A$ then satisfy (abbreviating $\ket{\pi(\vb{p})}$ to $\ket{\vb{p}}$)
\begin{widetext}
\begin{equation}
    \label{main-result}
    \begin{split}
        \Delta \rho_A\qty(\{\vb{x}\},\{\vb{x}'\}) = \rho_A\qty(\{\vb{x}\},\{\vb{x}'\})  \int \frac{\dd^3 p}{(2\pi)^3} ~\rho_\pi(\vb{p})~ & \bigg(\tfrac{\iu}{2} \bra{\vb{p}}\big( T_{\{\vb{x}\}}^{}+T_{\{\vb{x}\}}^\dagger -T_{\{\vb{x}'\}}^{}-T_{\{\vb{x}'\}}^\dagger\big)\ket{\vb{p}}   \\
        &\hspace{0.05\linewidth}- \tfrac12\bra{\vb{p}}\qty(   T_{\{\vb{x}\}}^{} T_{\{\vb{x}\}}^\dagger+  T_{\{\vb{x}'\}}^{}T_{\{\vb{x}'\}}^\dagger)\ket{\vb{p}} \\
        &
        \hspace{0.235\linewidth} +  \bra{\vb{p}} T_{\{\vb{x}'\}}^\dagger T_{\{\vb{x}\}}^{} \ket{\vb{p}} \bigg)~.
    \end{split}
\end{equation}
\end{widetext}
We stress that \cref{main-result} {\it does not} assume that atoms are localized in position eigenstates and does not rely on the Born approximation. Rather, we have expanded a (completely general) density matrix in the coordinate representation, and used the (exact) $T_{\{\vb{x}\}}$ matrix for the probe states for each configuration of atoms in the distribution.

\Cref{main-result} is our major result, and what follows are simple applications of this formula.
Notice that \cref{main-result} is manifestly traceless
and Hermitian and, by the invariance of $T_{\{\vb{x}\}}$ under particle relabeling, 
there also exist off-diagonal entries in \cref{main-result} that vanish. 
The first line of \cref{main-result} corresponds to the forward scattering phase, and contains non-trivial contributions starting at $O(V^2)$.\!\footnote{Since the probe momentum in- and out- states are identical, the first line of \cref{main-result} is simply given by \cref{eq:T-matrix-M}, where the scattering matrix element is evaluated for $\mathbf{p}=\mathbf{p'}$; at first order in the Born series, $\mathcal{M}^{(1)}_{\{\mathbf{x}\}}(\mathbf{p},\mathbf{p}) = \sum_{i} \widetilde{V}(\mathbf{0})$, which implies that the first line of \cref{main-result} vanishes at this order.} The second- and third-lines, which correspond to the decoherence part of the Lindblad equation, contain contributions starting at second order in the Born series, since $T$ contains contributions starting at $O(V)$. 
Whenever $T_{\{\vb{x}\}} \neq T_{\{\vb{x}'\}} $ the forward scattering phase is observable and we comment on its impact on specific measurements in what follows. 

\Cref{main-result} reduces to the standard result for a single atom~\cite{Gallis:1990,Hornberger:2003}. The first line has only forward scattering amplitudes and vanishes for a single atom, as shown in Ref.~\cite{Hornberger:2003}.  However, it does not vanish for $N\geq2$ and therefore represents a phase shift which is a {\it bona fide} $N$-particle effect, and which can survive even for an isotropic density matrix for probe states. 

Interpreting each unitary $S$-matrix as inducing a change in the state over a small interval of time $\Delta t$, the above equations can be re-written in differential form, 
\begin{equation}
    \label{eq:eom-Nbody}
    \dv{t} \rho_A(\{\vb{x}\}, \{\vb{x}'\}) =  -\lambda(\{\vb{x}\},\{\vb{x}'\})\rho_A(\{\vb{x}\},\{\vb{x}'\}).
\end{equation}
The solution of this differential equation is trivial being given by
\begin{equation}
\begin{aligned}
    \label{rho-evol}
    \rho_A(\{\vb{x}\},\{\vb{x}'\}, t) & = \rho_A(\{\vb{x}\},\{\vb{x}'\}) \\ & \, \times \exp\qty[ -\int_0^t \dd \tau \lambda(\{\vb{x}\},\{\vb{x}'\},\tau)] \, .
\end{aligned}
\end{equation}
We have anticipated the form of the evolution equation in a semi-classical picture where atomic positions change as a function of time $\vb{x} \rightarrow \vb{x}(\tau)$. For example if $\hat{H}_A$ is taken to be the free Hamiltonian then $\vb{x}(\tau) = \vb{x}_0 + \vb{v}_0 \tau$.  
The function $\lambda$ is calculable and can be decomposed into a unitary, $\lambda_U$, and ``decohering'' component, $\lambda_D$, which are identified by their appearance in the first line and subsequent lines of \cref{main-result}, respectively. 
 At leading order in the Born series, the expressions for $\lambda_U$ and $\lambda_D$ are given via \cref{eq:first-order-m-element,eq:second-order-m-element} by 
\begin{widetext}
\begin{equation}
\begin{aligned}
     \lambda_U &=  \int \frac{\dd^3 p}{(2\pi)^3}  ~\rho_\pi(\vb{p}) \int \frac{\dd^3q'}{(2\pi)^3} ~\widetilde{V} (-\vb{q'}) \qty[G_\pi(\vb{p}+\vb{q'})+ G_\pi^\dagger(\vb{p}+\vb{q'})]\widetilde{V} (\vb{q'}) \bigg[\frac{\iu}{2} \, \sum_{ij}^{N}\e^{\iu \vb{q'}\cdot(\vb{x}_i' -\vb{x}_j')}-\e^{\iu \vb{q'}\cdot(\vb{x}_i -\vb{x}_j)}\bigg]~,
\end{aligned}     
\label{lambda-u}
\end{equation}
\begin{equation}
\begin{aligned}
    \label{lambda-d}
    \lambda_D &= \int \frac{\dd^3 p}{(2\pi)^3} ~\rho_\pi(\vb{p})\int\frac{\dd^3q}{(2\pi)^3}~ (2\pi) \delta(\Sigma E) |\widetilde{V} (\vb{q})|^2  \bigg[\frac12\sum_{ij}^{N} \e^{\iu \vb{q}\cdot(\vb{x}_i -\vb{x}_j)} + \e^{\iu \vb{q}\cdot(\vb{x}'_i -\vb{x}'_j)}- 2\e^{\iu \vb{q}\cdot(\vb{x}_i -\vb{x}_j')}  \bigg] ~,
\end{aligned}
\end{equation}
\end{widetext}
where $\vb{q'}$ is not to be confused with the momentum transferred by the probe.
\Cref{lambda-d,lambda-u} factorize into a real function $\omega_{U,D}$ that depends solely on the kinematics [i.e.,~after averaging over $\rho_\pi(\vb{p})$], and a kernel (in square brackets) $K_{U,D}(\{\vb{x}\},\{\vb{x'}\})$.
For definiteness, in what follows we will treat the $N$ particles as distinguishable; this is justified for atomic de Broglie wavelengths that are short relative to the interparticle spacing.  


In the following, we aim to understand the way in which coherent (multi-atom) effects differentiate the imprints of the environment on an experiment employing a $N$-atomic cloud from the imprints of the environment on $N$ iterations of an experiment employing a single atom.

\section{Toy model of a two-arm interferometer \label{sec:Two-Arm}} 
 The parametric scaling of \cref{lambda-d,lambda-u} with $N$ can be most easily understood in a two-mode interferometer with $N$ atoms, for which the Hilbert space is $2^N$-dimensional. In this model, the position labels $\{ \vb{x}\}$ and $\{ \vb{x}'\}$ assume discrete values which we label as $L$ or $R$,  e.g., $\{\vb{x}\} = \{L,L,R,L,L,R\}$ for a specific configuration of a six-atom system. The two discrete positions $L$ and $R$ are separated by a (time dependent) distance $\Delta\vb{x} = | \vb{x}_L-\vb{x}_R|$.  This approximates a realistic two-arm interferometer when momenta transfers are sufficiently small that the cloud appears point-like (i.e., it corresponds to the limit where the cloud radius is taken to be much smaller than the typical inverse momentum transfer). 
 We will label the number of atoms in the left arm of $\ket{\{\vb{x}\}}$ by $N_L$ (e.g., $N_L =4$ for $\ket{\{LLRLLR\}}$) and the number of atoms in the left arm of $\bra{\{\vb{x}'\}}$ by $N_L'$, and define $n = N_L-N_L'$ as the atom asymmetry. With this convention, in the limit of coherent scattering $|\mathbf{q}| < 1/ r_c$, being $r_c$ the cloud radius, we may evaluate the kernels, defined as the terms in between square brackets in \cref{lambda-u} -- unitary kernel-- and \cref{lambda-d} --decoherence kernel, for any $N$-tuple $\{\vb{x}\}$ and $\{\vb{x'}\}$
\begin{align}
    \label{Decoherence-kernel-u-toy-model}
     K_U &= \iu \, n (n + N - 2N_L) \qty[1-\cos(\vb{q'}\cdot \Delta \vb{x})]  ~,\\
     K_D &= n^2 \qty[1-\cos(\vb{q}\cdot \Delta \vb{x})] - \iu N n \sin(\vb{q}\cdot \Delta \vb{x}) \, , 
     \label{Decoherence-kernel-d-toy-model}
\end{align}
where $-N \leq n \leq N$ and $0 \leq N_L \leq N$, and $|\Delta \vb{x}|$ is the separation between the interferometer arms (\textit{cf.} Appendix~\ref{app:combinatorics} for a derivation of these kernel formulas). 
Notice that both kernels are invariant under permutations of the atoms' labels and therefore vanish for $n=0$. 
This also agrees with the information-theoretic understanding of decoherence~\cite{Schlosshauer:2019ewh}:
the larger the asymmetry between states, the more ``which-path'' information is gathered by a single scattering event.
Pairs of $N$-particle states for which $n=0$ (e.g., $\ket{LR}\bra{RL}$ for $N=2$ or $\ket{LLR}\bra{LRL}$ for $N=3$) are invariant under open-system dynamics induced by the probe-system scattering process and
are therefore elements of decoherence-free subspaces~\cite{Lidar:2014xxx}. We note that $K_D \to 0$ in the limit where the probe particle does not resolve the two interferometer arms, $|\mathbf{q}| \ll 1/|\mathbf{\Delta x}|$, as expected from a probe particle (potential observer) that does not localize where the atoms are amongst the two interferometer paths.

\subsection{Product states \label{sec:Product-States}} 
Let us now consider a two-arm interferometer in which the atoms are initially uncorrelated and each prepared in a superposition of the left and right arm,
$\frac{1}{\sqrt{2}}(\ket{L} + \ket{R})$. This many-particle wavefunction can be written as a product state $\ket{\Psi}= \bigotimes_{i=1}^N \frac{1}{\sqrt{2}}(\ket{L_i} + \ket{R_i})$. This may be considered as a toy model of an atom interferometer, such as the 10-meter atom fountain at  Stanford~\cite{Asenbaum:2020era}, or the proposed MAGIS~\cite{Abe:2021} and AION~\cite{Badurina:2019hst} experiments (whose primary goals are to search for mid-frequency gravitational waves and ultralight dark matter~\cite{Badurina:2021rgt}), employing a dilute atom cloud of radius $r_c\ll |\Delta \vb{x}|$ and $r_c\ll 1/|\vb{q}|$,  where $|\mathbf{q}|$ is the momentum transferred by the probe.
In this case the density matrix is not sparse, and one must consider all entries in $\rho_A$, for which the decoherence kernel is given by \cref{Decoherence-kernel-d-toy-model}.

\Cref{Decoherence-kernel-d-toy-model} shows that 
density matrix elements with $n \sim O(N)$ have coherently enhanced rates of decoherence. 
A typical measurement in an atom interferometer, however, is a one-body measurement~\cite{abend2020atom}, i.e., the observable is represented by an operator $\mathcal{O}_{\rm 1B} =\sum_{i=1}^N \mathcal{O}_i$ where $i$ labels each atom in the gas. In this case, because $N-1$ single-atom Hilbert spaces are traced over, only terms  
with $|n|\leq 1$ in $\rho_A$ contribute to the expectation value of the observable.

For example, in atom interferometers based on diffuse atomic clouds, the accessible experimental observable is the relative number of atoms measured in a given port, e.g., $|+\rangle$, with respect to the total number of atoms measured. A fringe is typically inferred using $\mathcal{O}_i = \ket{+_i}\bra{+_i}$ with $\ket{+_i}=\frac{1}{\sqrt{2}}(\ket{L_i}+\ket{R_i})$ \cite{abend2020atom,Geiger:2020aeq} (which projects the atoms in port $|+\rangle$). 
This fringe is characterized by its amplitude, usually named visibility/contrast $V$ and phase shift $\varphi$.
The expectation value of $\mathcal{O}_+=\sum_{i=1}^N \mathcal{O}_i$ is then given by 
\begin{equation}\label{eq:expectation_O+}
    \left\langle \mathcal{O}_+ \right \rangle = \mathrm{Tr}(\rho_A \mathcal{O}_+) = 
    \tfrac{N}{2} \big(1+ V \cos\varphi \big)~,
\end{equation}
where the trace is over the $2^N$-dimensional Hilbert space. The measured contrast and phase shift are related to the one-body reduced density matrix $\rho_1={\rm Tr}_{N-1} (\rho_A$) via 
\begin{equation}\label{eq:vis_phase}
    V \cos \varphi = 2 \, \mathrm{Re}(\mel{L}{\rho_{1}}{R}) \, .
\end{equation}
After accumulating a dynamical (i.e., independent of open-quantum system dynamics) phase-shift $\phi$, the off-diagonal element of $\rho_1$ can be written as
\begin{equation}
    \label{rho1-expl}
        \mel{L}{\rho_{1}}{R} = \frac12\cos^{N-1}(\tau)~\e^{-s+\iu\qty[ \phi-N\gamma]} \, ,
\end{equation}
where,
\begin{align}\label{eq:s}
    s&=\int \dd t \int \frac{\dd^3q}{(2\pi)^3}  ~~  \omega_D(\mathbf{q},t) \qty[1-\cos(\vb{q}\cdot \Delta \vb{x})]~,\\ \label{eq:gamma}
    \gamma&=\int \dd t \int \frac{\dd^3q}{(2\pi)^3} ~~ \omega_D(\mathbf{q},t) \sin(\vb{q}\cdot \Delta \vb{x})~,\\
    \tau &=\int \dd t \int \frac{\dd^3q'}{(2\pi)^3}~~   \omega_U(\mathbf{q'},t)\qty[1-\cos(\vb{q'}\cdot \Delta \vb{x})]~, \label{eq:tau}
\end{align}
are real functions which scale as $|\widetilde{V} (\vb{q})|^2$ and parameterize previously-known ($s$ and $\gamma$) \cite{Gallis:1990,Hornberger:2003} and novel, ($\tau$) collision-induced effects. We refer the reader to Appendix~\ref{app:reduced-rho} for the derivation of a generic matrix element of a $\alpha$-body density matrix (the results above can be obtained by taking $\alpha = 1$, ${\cal N}_L = 1$ and ${\cal N}_L'=0$ in Eq.~\eqref{app:eq:short-hand}). The phase $\tau$, which originates from forward scattering, is only observable for $N\geq 2$ and does not vanish even for a rotationally invariant $\rho_\pi(\vb{p})$ (e.g.~a Maxwell-Boltzmann gas).\footnote{For a discussion on the non-vanishing forward scattering phase, see \cref{app:forward-scatt}.} The cosine in \cref{rho1-expl} arises from (destructive) interference between different terms in the partial trace. Both $\gamma$ and $\tau$ receive $N$-enhancements while $s$ does not, in agreement with \cref{Decoherence-kernel-d-toy-model,Decoherence-kernel-u-toy-model} for $|n|=1$. Although the overall amount of decoherence, $\e^{-s}$, is not enhanced by factors of $N$, there can still be a loss of contrast from dephasing, $\cos^{N-1}(\tau)$, which is both enhanced in the large-$N$ limit and absent in the single atom case. Therefore, $\tau$ can be measured using the standard contrast/visibility observable.
Furthermore, in the regime where the probe particle does not resolve the two interferometer paths, $|\mathbf{q}| \ll 1/|\mathbf{\Delta x}|$, the integrands of $s$ 
and $\gamma$ are suppressed by $(\mathbf{q}\cdot \mathbf{\Delta x})^2$ and $(\mathbf{q}\cdot \mathbf{\Delta x})$, respectively. 
However, the unitary kernel in \cref{Decoherence-kernel-u-toy-model} still contributes to the contrast loss via the phase $\tau$.

\subsection{Statistical fluctuations of one-body operators \label{sec:Stat-Fluct}} 
Higher-order moments of one-body operators offer a natural probe of enhanced  decoherence rates. For example consider counting atoms in a port ``$+$''. After $\nMeas$ runs of the experiment the collected data can be written as the sequence $\{ n_+^{(1)}, n_+^{(2)}, n_+^{(3)}, \ldots, n_+^{(\nMeas)}\}$. The fluctuations of $n_+^{(i)}$ are sensitive to enhancements that scale like $n^2$. As a simple illustration let us consider the variance (second moment) estimated using  $\frac{1}{\nMeas}\sum_{a=1}^\nMeas [n_+^{(a)}  - \langle n_+\rangle]^2$, where $\langle n_+\rangle=\frac{1}{\nMeas}\sum_{a=1}^\nMeas n_+^{(a)}$ is the estimator of the mean. 

The counting operation corresponds to the one-body operator $\mathcal{O}_{+}$ previously defined. The mean number of counts in this port is $\langle \mathcal{O}_+ \rangle = {\rm Tr}\qty[\rho \, \mathcal{O}_{+} ]$, while the variance is given by $\sigma_{+}^2=  \langle \mathcal{O}_{+}^2 \rangle - \langle \mathcal{O}_{+} \rangle^2 = {\rm Tr}\qty[\rho \, \mathcal{O}_{+}^2] - \langle \mathcal{O}_+ \rangle^2$. The two-body operator $\mathcal{O}_+^2$ can be expanded as
\begin{equation}
    \label{OA-sq}
    \mathcal{O}_{+}^2 =  \sum_{i=1}^N \sum_{j\neq i}^N 
    \mathcal{O}_i\mathcal{O}_j
    ~+~\sum_{i=1}^N \mathcal{O}_i~.
\end{equation} 
To evaluate $\langle \mathcal{O}_{+}^2 \rangle$, one must use $\rho_2={\rm Tr}_{N-2}\rho_A$ see e.g., Ref.~\cite{PhysRevA.47.3554}. The two-body reduced density matrix, $\rho_2$, will have two entries which decohere with a rate enhanced by $(n=2)^2$. We show this schematically below for the elements of $\rho_2$ labeled (left to right, and top to bottom) by states $\ket{LL}$, $\ket{LR}$, $\ket{RL}$, and $\ket{RR}$
\begin{equation}
    \label{matrix}
   \rho_2 = 
   \begin{pmatrix}
        \circ ~       & ~\blacksquare~ & ~\blacksquare~  & ~\bigstar \\
        \blacksquare~ & ~\circ~        & ~\circ~         & ~\blacksquare \\
        \blacksquare~ & ~\circ~        &  ~\circ~        & ~\blacksquare \\
        \bigstar ~    & ~\blacksquare~ & ~\blacksquare~  & ~\circ
    \end{pmatrix}~.
\end{equation}
Entries with $\circ$ have $n=0$, entries with $\blacksquare$ have $|n|=1$, and entries with $\bigstar$ have $n=2$.  The variance ${\sigma_+^2 = \langle \mathcal{O}_+^2 \rangle - \langle \mathcal{O}_+ \rangle^2}$ can be constructed using
\begin{equation}
\begin{aligned}
    \label{matrix-in-math}
    \langle \mathcal{O}_+^2\rangle = \langle \mathcal{O}_+ \rangle + \frac{N(N-1)}{4}
     & \bigg[\frac{3}{2}+ 4{\rm Re} \mel{L}{\rho_1}{R} \\
    & \, \, \,  + 2{\rm Re}\mel{LL}{\rho_2}{RR}\bigg]~.
\end{aligned}
\end{equation}
The corner-entry of $\rho_2$ may be written as 
\begin{equation}
    \begin{split}
        \label{rho2-expl}
       \mel{LL}{\rho_2}{RR} = &\frac14\cos^{N-2}(2\tau) ~\e^{-4s+2\iu \qty[\phi-N\gamma ]}~,
    \end{split}
\end{equation}
for $N\geq2$.
\Cref{rho2-expl} exhibits enhanced decoherence relative to \cref{rho1-expl} due to $\mel{LL}{\rho_2}{RR}$ having $n=2$. More generally, the $\eta^{\rm th}$ moment of $\mathcal{O}_+$ is computed using $\rho_\eta={\rm Tr}_{N-\eta}\rho_A$ whose corner entry has $n=\eta$ (\textit{cf.} \cref{app:reduced-rho} for a discussion on reduced density matrices and \cref{app:stat_fluct} for general expressions
of the statistical fluctuations of one-body operators). Thus, for $\eta \sim O(N)$ one can construct observables which experience rates of decoherence enhanced by $O(N)$.

Examining \cref{matrix,matrix-in-math} one finds another interesting phenomenon that occurs in the limit of $\e^{-s}\rightarrow 0$. Although the mean follows what would be expected from a conventional classical state of coin flipping, $\langle\mathcal{O}_+\rangle = 1/2$, we find that the variance remarkably scales as $N^2$ instead of $N$,
\begin{equation}
    \label{funny-stats}
    \sigma^2_+ = \frac{N(N+1)}{8}~.
\end{equation}
The statistics follow a super-Poissonian distribution. 
This is a consequence of the non-zero off-diagonal entries ($\circ$) with $n=0$ in \cref{matrix} which  would vanish for a maximally mixed state. These entries do not decohere because of the permutation symmetry of the atomic positions $\{\vb{x}\}$ and $ \{\vb{x}'\}$, {\it cf}. \cref{main-result}. Although this result can be anticipated mathematically, it is classically counterintuitive. For ultra feeble interactions, such as those of dark matter, the $\e^{-s} \rightarrow 0$ limit will not occur. However, for standard environmental decoherence this limit is ubiquitous. \Cref{funny-stats} may therefore be interpreted as a generic signal for a system which has been strongly decohered as a result of coherent scattering. 

\subsection{Entangled states \label{sec:Entangled}} 
Before concluding, let us consider an example that illustrates the $N^2$-enhancements, and that is simple to analyze theoretically. Suppose that the $N$-atom system is prepared in the maximally-entangled ($N00N$) state ${\ket{\Psi}=\frac{1}{\sqrt{2}}(\bigotimes_{i=1}^N \ket{L_i} + \bigotimes_{i=1}^N \ket{R_i})}$. 
This configuration mimics a matter-wave interferometer, which is similarly entangled but due to intra-material forces rather than state preparation. 
For such a system preparation, there are only two non-vanishing off-diagonal entries in $\rho_{A}$. 
These correspond to $(n, N_L)=(N,N)$ and $(n, N_L)=(-N,0)$. 
Using \cref{Decoherence-kernel-u-toy-model,Decoherence-kernel-d-toy-model} derived in the limit of coherent scattering $1/|\mathbf{q}| < r_c$, one finds $K_U=0$ and 
\begin{align}
    \label{N00N-kernel}
    K_D(\vb{q},\{\vb{x}\},\{\vb{x}'\}) = N^2 (\e^{\pm \iu \vb{q}\cdot(\vb{x}-\vb{x}')}-1)~,
\end{align}
equivalent to the replacement $|\widetilde{V} (\vb{q})|^2 \rightarrow N^2 |\widetilde{V} (\vb{q})|^2$ in the single atom formula.  An atom interferometer employing $N00N$ states and measuring $\mathcal{O} = \ket{\psi}\bra{\psi}$ is therefore susceptible to coherent enhancements.

$N00N$ states with $N\gtrsim 10$ are difficult
systems to prepare in the lab; 
hence, we do not expect $N00N$ states themselves to play an important role in future atom interferometer searches for dark matter. 
Nevertheless, the parametric scaling we have identified applies to other systems of experimental relevance. We have in mind, in particular, matter-wave interferometers employing mesoscopic objects. Ordinary matter naturally contains both a large number of particles and a high level of entanglement. The state of, e.g., a piece of gold, can be described by a many-body wavefunction in which the relative coordinates are tightly-localized (and therefore entangled) about the center of mass coordinate. In a matter-wave interferometer, the center of mass can  be delocalized, and the system placed in a quantum superposition with highly entangled relative coordinates. This situation is closely analogous to the $N00N$ state analyzed above. \Cref{N00N-kernel} suggests that existing approaches in the literature~\cite{Riedel:2012ur, Riedel:2016acj, Du:2022ceh, Du:2023eae} for computing decoherence for \textit{matter-wave} interferometers, such as MAQRO~\cite{Kaltenbaek:2015kha}, are therefore reliable. While these matter interferometers offer experimental challenges of their own right, they offer a clear path towards macroscopic atom populations i.e., $N \sim 10^{10}$ for the MAQRO experiment~\cite{Kaltenbaek:2015kha}. 

\section{Discussion and conclusions \label{sec:Disc-Concl}} 
In this work we have computed the collisional decoherence of a system of $N$ atoms due to scattering with a background gas. Specifically, \cref{main-result} can be used to compute the collisional decoherence of a generic $N$-particle system. Focusing on the case when the probe states scatter coherently off of the atoms, we find that a careful treatment of the $N$-body system is required to properly characterize both the phase shift and loss of visibility arising from collisional decoherence {\it cf}.~\cref{eq:expectation_O+}. Indeed, there exist inherently $(N\geq 2)$-body effects that can be searched for in future two-arm interferometers. Importantly, the observation of this phase shift and of coherently-enhanced decoherence rates depends on: {\it i)} the initial state preparation, {\it ii)} the dynamics of the interferometer, and {\it iii)} the final measurement being performed. 

Although we have focused on a toy-model of an interferometer, our qualitative conclusions have immediate consequences for dark matter direct detection ~\cite{Riedel:2012ur,Riedel:2016acj,Du:2022ceh,Du:2023eae}, and atom interferometers more generally.  While we find that one-body observables are less sensitive to coherent enhancements than one would naively expect, it is clear that {\it some} coherent effects can be used as a resource for the discovery of dark matter. For example atom interferometers employing $N00N$ states and matter-wave interferometers naturally exhibit coherently enhanced decoherence. Similarly, statistical fluctuations of counting measurements in atom interferometers offer a probe of coherent enhancements using uncorrelated initial states. We leave a more detailed investigation into optimal strategies for measuring decoherence and modeling of proposed and existing experimental configurations to future work.  Nevertheless we have identified observable $N$-enhancements for: the phase $\gamma$, loss of contrast from dephasing due to the forward scattering phase $\tau$, and decoherence that enters from higher statistical moments of one-body observables. All of these effects are resources that enhance sensitivity when searching for decoherence from feebly interacting environmental particles such as light dark matter.

It would be interesting to test the features of decoherence that are inherent to ($N\geq 2$)-particle systems in the lab.  This could be achieved even for a system with a modest number of atoms (e.g., $N\sim 10$) and a background gas which has a large coherent cross section with atoms in the limit of low momentum transfer. One could study higher order moments of single-body operators as described above, and search for coherently enhanced rates of decoherence and/or the influence of the forward scattering phase. Ideally, to match onto the two-mode approximation employed in this work, the radial size $r_c$ of the $N$-atom system (not to be confused with the separation between the arms of the interferometer) would have to satisfy $r_c \ll 1/|\vb{q}_{\rm max}|$ with $|\vb{q}_{\rm max}| \approx \sqrt{2 m T}$ with $T$ and $m$ the temperature and mass of the background gas, respectively. Furthermore, while \cref{main-result} applies generally to any $N$-atom system, the expressions in \cref{lambda-u,lambda-d} rely on the Born approximation. Therefore, if these analytic results are to be used, then one should have a dilute gas satisfying $na^3\ll1$ where $n$ is the number density of the atoms and $a$ is the atom-atom scattering length \cite{abrikosov2012methods,Braaten:1996rq,Hammer:2000xg}. 
We leave a more detailed modeling of effects (e.g., system substructure, finite temperature, etc.) that go beyond the idealized limit discussed above to future work. 

The formalism presented here can be tested in future experiments, including searches for the forward scattering phase, and coherently enhanced rates of decoherence. This would both shed light on the quantum to classical transition, and aid in the design and optimization of future interferometers.  The forward scattering phase, $\tau$, and the decoherence phase, $\gamma$, are enhanced by $O(N)$, and may be an important background (or new signal) when searching for anomalous phase shifts in atom interferometers. Theoretical control over coherently enhanced effects will, in turn, benefit dark matter searches, gravitational wave detection, and other applications to fundamental physics (see Refs.~\cite{Chou:2023hcc, Proceedings:2023mkp} and references therein).

\section*{Acknowledgements}
We thank Klaus Hornberger, Duncan O'Dell and John E.~Sipe for feedback during the completion of this work. We also thank Daniel Carney for discussions on the validity of the Born approximation and comments on an early version of this paper. We are especially grateful to Kathryn M.~Zurek for many stimulating discussions and comments on the manuscript.

C.M.~is supported by ``Ayuda Beatriz Galindo Junior'' from the Spanish ``Ministerio de Universidades'', grant BG22/00155. C.M.~thanks Caltech for its hospitality during the completion of this work. R.P. and L.B.~are supported by the U.S. Department of Energy, Office of Science, Office of High Energy Physics under Award Number DE-SC0011632, and by the Walter Burke Institute for Theoretical Physics.  R.P.~is supported by the Neutrino Theory Network under Award Number DEAC02-07CH11359.

\begin{appendix}

\section{Scattering theory conventions}\label{app:scatt_conventions}
For a self-contained discussion, we provide here our conventions for the normalization of states and the definition of the scattering cross section. We take our states to be relativistically normalized 
\begin{equation}
    \braket{p'}{p}= 2 E_p (2\pi)^3 \delta^{(3)}(\vb{p}-\vb{p}')~. 
\end{equation}
Scattering matrix elements are defined in terms of the $T$-matrix, $S=1+\iu T$, 
\begin{equation}
    \mel{f}{T}{i} = (2\pi)^4 \delta^{(4)}(\Sigma p) \mathcal{M}_{fi}~,    
\end{equation}
where $\Sigma p = \Sigma p_f -\Sigma p_i$, and $p_{i,f}$ are the four momenta of the initial and final states, respectively.
Scattering cross sections involving $2 \rightarrow n$ processes are given by 
\begin{equation}
\sigma_{fi} = \frac{1}{\mathcal{F}}\! \int \!\dd \Pi_1 \ldots\! \dd \Pi_n   (2\pi)^4 \delta^{(4)}(\Sigma p) | \mathcal{M}_{fi} |^2~,
\end{equation}
where $\dd \Pi = \dd^3 p/(2E_p)/(2\pi)^3$, and $\mathcal{F}=2\sqrt{\lambda(s,m_1^2,m_2^2)}$ is a Lorentz invariant flux normalization with $\lambda(x,y,z)= x^2+y^2+z^2-2xy -2yz -2xz$ the K\"allen triangle function. The masses of the incoming particles are $m_1$ and $m_2$. Using the above expression it is straightforward to derive 
\begin{equation}
   \dv{\sigma}{ \Omega} =  \int\frac{\dd^3q}{(2\pi)^3} (2\pi) \delta(\Sigma E) |\mathcal{M}|^2~, 
\end{equation}
where $\Sigma E = \Sigma E_f -\Sigma E_i$, for $2\rightarrow 2$ scattering in the static limit where $s-m_1^2 \ll m_1^2$. 

It is also useful to introduce the relevant normalization for potential scattering. These are frame dependent since the static background field implicitly selects a preferred frame, i.e.,~the rest frame of the target. Scattering in a background field is a $1\rightarrow 1$ process and in this case the normalization of $T$-matrix elements is 
\begin{equation}
    \mel{p'}{T}{p} = (2\pi) \delta(E' - E) \mathcal{M}(p',p)~.
\end{equation}
The differential cross section is written in terms of the potential-scattering matrix element as 
\begin{equation}
    \sigma v = \frac{1}{2 E} \int \dd \Pi  ~ (2\pi) \delta(\Sigma E) ~ |\mathcal{M}|^2~, 
\end{equation}
where $v=|\vb{p}|/E$ is the velocity of the incident particle in the laboratory frame. When scattering from a background potential, the matrix element at first order in the Born approximation is related to the Fourier transform of the potential $\mathcal{M} = (2 E) \widetilde{V}(\vb{q})$. The factor of $(2E)$ is related to the relativistic normalization of states.   

\section{Potential scattering derivation}\label{app:pot_scatt}
In this section we explain how the background potential approximation emerges from the analysis of Feynman diagrams. This helps resolve certain conceptual ambiguities. For example, if one considers $3\rightarrow 3$ scattering involving atom $i$, atom $j$, and the probe $\pi$, then the associated Feynman diagram appears as a tree-level graph. However, in the potential scattering calculation, this diagram is second order in the Born series, and one must integrate over momentum transfers. This ``paradox'' arises because we consider atoms in position eigenstates, rather than momentum eigenstates. We discuss this point in detail in what follows. 

Let us work out the scattering of a probe on a single particle at a fixed position,
\begin{equation}
    \ket{\vb{x}} = \int \frac{\dd^3 k}{(2\pi)^3} \e^{\iu \vb{k} \cdot \vb{x}} \ket{\vb{k}}~.
\end{equation}
As a short-hand we will use $[\dd k] = \dd^3 k/(2\pi)^3$. We are interested in 
\begin{equation}
     \bra{\vb{p}'}T\ket{\vb{x},\vb{p}} = \int[\dd k] \e^{\iu \vb{k} \cdot \vb{x}} \bra{\vb{p}'}T\ket{\vb{k},\vb{p}}~,
\end{equation}
where $T$ is the $T$-matrix appearing in the definition of the $S$-matrix, i.e., $S = 1 + \iu \, T$.
Inserting a complete set of atomic momentum eigenstates, 
\begin{equation}
     \bra{\vb{p}'}T\ket{\vb{x},\vb{p}} = \int [ \dd k]  [ \dd k'] \e^{\iu \vb{k} \cdot \vb{x}} \ket{\vb{k}'}\bra{\vb{p}',\vb{k}'}T\ket{\vb{k},\vb{p}}~.
\end{equation}
Using the definition of the scattering matrix element
\begin{equation}\label{app:eq:step}
    \begin{split}
    \bra{\vb{p}',\vb{k}'}T\ket{\vb{k},\vb{p}}
    & = (2\pi)^4 \delta^{(4)}(\Sigma p) ~\iu\mathcal{M}(v\cdot q,q^2)~,
    \end{split}
\end{equation}
where we assume that the matrix element depends only on the Lorentz invariants $v\cdot q$ and $q^2$, with $v$ the atom's four velocity\footnote{This approximation is related to the hierarchy of scales $|\vb{q}| \ll M_A$. For an atom at rest, $v_\mu = (1,0,0,0)$.}, and $q_\mu=p'_\mu-p_\mu$. To first order (i.e., the probe scattering off of an atom once), the corresponding Feynman diagram for the $i$-th atom is (trivially), 
\vspace{6pt}
\begin{equation}
\begin{gathered}
\begin{fmffile}{diagram_soft}
 \fmfstraight
\begin{fmfgraph*}(20,7) 
    \fmfleft{i1,i2}
    \fmfright{o1,o2}
    \fmf{double}{i1,b1,o1}
    \fmf{plain}{i2,t1,o2}
    \fmf{dashes,tension=0,lab=$\vb{q}$}{b1,t1}
    \fmflabel{$\pi$}{i2}
    \fmflabel{$\vb{x}_i$}{i1}
\end{fmfgraph*}
\end{fmffile}  
\end{gathered}
\end{equation}
At second order mediators can talk to the same atom twice. For the $i$-th atom, this corresponds to a two-body one-loop graph,
\vspace{6pt}
\begin{equation}\label{app:eq:loop}
\begin{gathered}
\begin{fmffile}{diagram_soft_1}
 \fmfstraight
\begin{fmfgraph*}(20,9) 
    \fmfleft{i1,i2,fi,i3}
    \fmfright{o1,o2,fo,o3}
    \fmf{double}{i2,b1,b2,o2}
    \fmf{plain}{i3,t1,t2,o3}
    \fmf{dashes,tension=0,lab=$\vb{q}_1$}{b1,t1}
    \fmf{dashes,tension=0,lab=$\vb{q}_2$}{b2,t2}
    \fmflabel{$\pi$}{i3}
    \fmflabel{$\vb{x}_i$}{i2}
\end{fmfgraph*}
\end{fmffile}   
\end{gathered}
\end{equation}
The two momentum transfers $\vb{q_1}$ and $\vb{q_2}$ are constrained by $\vb{q_1}+\vb{q_2}=\vb{q}$. Using the delta function to integrate over $\vb{k'}$, \cref{app:eq:step} takes the form 
\begin{equation}
    \begin{split}
    \bra{\vb{p}'}T\ket{\vb{x},\vb{p}} =& \int [ \dd k]   \e^{\iu \vb{k} \cdot \vb{x}}\ket{\vb{k}+\vb{q}}
    (2\pi) \delta(\Sigma E) \iu\mathcal{M}(v\cdot q,q^2)~.
    \end{split}
\end{equation}
Next, writing $\ket{\vb{k}+\vb{q} }= \e^{\iu \vb{q} \cdot \hat{\vb{x}}}\ket{\vb{k}}$, and then using the fact that the matrix element and energy conserving delta function do not depend on $\vb{k}$, we find, 
\begin{equation}
    \begin{split}
    \bra{\vb{p}'}T\ket{\vb{x},\vb{p}} &=  (2\pi) \delta(\Sigma E) \iu\mathcal{M}(v\cdot q,q^2) \e^{\iu \vb{q} \cdot \hat{\vb{x}}}\ket{\vb{x}} \\
    &= (2\pi) \delta(\Sigma E) \iu\mathcal{M}(v\cdot q,q^2) \e^{\iu \vb{q} \cdot \vb{x} }\ket{\vb{x}}~,
    \end{split}
\end{equation}
where in the last equality we have acted with the position operator on the position eigenstate. We may re-write this as 
\begin{equation}
    T\left (\ket{\vb{x}}\otimes\ket{\vb{p}} \right ) = \ket{\vb{x}} \otimes( T_{\vb{x}} \ket{\vb{p}})~.
\end{equation}
Notice the that the induced $T$-matrix is normalized as is appropriate for potential scattering \cite{Berestetskii:1982qgu}. 

What if the probe scatters off of two different atoms?  In this case, the problem at hand is $3\rightarrow 3$ scattering. 
The leading order Feynman diagram appears as a tree-level graph,
\vspace{6pt}
\begin{equation}\label{app:eq:tree-level}
\begin{gathered}
\begin{fmffile}{diagram_soft_2}
 \fmfstraight
\begin{fmfgraph*}(20,9) 
    \fmfleft{i1,i2,fi,i3}
    \fmfright{o1,o2,fo,o3}
    \fmf{double}{i1,b1,f1,o1}
    \fmf{double}{i2,f2,b2,o2}
    \fmf{plain}{i3,t1,t2,o3}
    \fmf{dashes,tension=0,lab=$\vb{q}_1$}{b1,t1}
    \fmf{dashes,tension=0,lab=$\vb{q}_2$}{b2,t2}
    \fmflabel{$\pi$}{i3}
    \fmflabel{$\vb{x}_i$}{i2}
    \fmflabel{$\vb{x}_j$}{i1}
\end{fmfgraph*}
\end{fmffile}
\end{gathered}
\vspace{6pt}
\end{equation}
where $i\neq j$. 
Performing the exact same analysis as above in terms of position eigenstates and $T$-matrices, one arrives at 
\begin{equation}\label{app:eq:T-3_scatt}
    \begin{split}
    \bra{\vb{p}'}T\ket{\vb{x}_1,\vb{x}_2,\vb{p}} =&   (2\pi) \delta(\Sigma E) \times \\
    &
    \int [\dd q_1] \iu \mathcal{M}(\vb{q}_1,\vb{q}) \\
    &\hspace{0.1\linewidth}
    \e^{\iu \vb{q}_1 \cdot \vb{x}_1} \e^{\iu (\vb{q}-\vb{q}_1)\cdot \vb{x}_2} 
    \ket{\vb{x}_1,\vb{x}_2}~.
    \end{split}
\end{equation}
It follows that $T\left (\ket{\vb{x}_1,\vb{x}_2}\otimes\ket{\vb{p}} \right ) = \ket{\vb{x}_1,\vb{x}_2} \otimes( T_{(\vb{x}_1,\vb{x}_2)} \ket{\vb{p}})$.
\cref{app:eq:T-3_scatt} is the same amplitude that would be obtained using potential scattering Feynman rules for two potentials located at $\vb{x}_1$ and $\vb{x}_2$, i.e., the second order term in the Born series. Note that the matrix element is normalized with a single energy conserving delta function, as is appropriate for potential scattering. 

For decoherence, in which one typically considers small momentum transfers, it is most efficient to formulate the problem in the language of potential scattering. The probe state may be relativistic or non-relativistic. Iteration generates higher loop order diagrams from potential scattering such that the induced $S$-matrix corresponds to potential scattering at arbitrary order in perturbation theory. 

We have focused on a non-interacting dilute gas, allowing us to neglect inter-atomic interactions to a good approximation. In the instance where interactions among particles are important, the Feynman diagram approach here can be used to incorporate interactions between atoms. Such a program is often considered in the context of nuclear scattering where one-pion exchange graphs generate two-body potentials~\cite{Bedaque:2002mn,Hammer:2019poc}. One could similarly consider an interacting gas of probe states via the same approach. 

\section{Non-vanishing forward scattering phase}\label{app:forward-scatt}
The forward scattering phase does not vanish upon averaging over all angular directions of the probe's momentum $\vb{p}$. 
To see this explicitly, consider the Feynman diagrams in \cref{app:eq:tree-level,app:eq:loop} for $\vb{q} = \vb{p'-p} = 0$.
Recall that we assume the clouds to be point-like, i.e., the distance between interferometer arms is large, but the interparticle separation is small.  
Any term with $\vb{x}_i$ and $\vb{x}_j$ in different arms of an interferometer can be dropped due to their rapidly oscillating exponentials.  
For the remaining terms, with $\vb{x}_i$ and $\vb{x}_j$ in the same arm, we can make the replacement $\e^{\iu \vb{q'}\cdot(\vb{x}_i-\vb{x}_j)}\rightarrow 1$ in the loop integral. 
We are then left with the integral, 
\begin{equation}
    \mathcal{I}(\vb{p})  = \int \frac{\dd^3\vb{q'}}{(2\pi)^3} \qty|\widetilde{V} (\vb{q'})|^2~\qty[G_\pi(\vb{p}+\vb{q'}) + G_\pi^\dagger(\vb{p}+\vb{q'})]~.
\end{equation}
Assuming $\widetilde{V} (\vb{q'})$ is rotationally invariant, there is no additional reference vector in the problem except for $\vb{p}$. 
This implies that the integral written above can depend only on $\vb{p}^2$, $\mathcal{I}(\vb{p})=\mathcal{I}(\vb{p}^2)$.
 Therefore the average over initial particle directions does not vanish even for isotropic probe momentum distributions, 
\begin{equation}
    \int [\dd p] ~\rho_\pi(\vb{p})~ \mathcal{I}(\vb{p}^2) \neq 0 ~.
\end{equation}

\section{Combinatorics}~\label{app:combinatorics}
In the limit of coherent scattering, the momenta transfers are sufficiently small that the cloud in each arm of the experimental appears point-like. Hence, we may assume that an atom in the left arm is located at position $\vb{x}_L$, while an atom in the right arm is located at position $\vb{x}_R$.
Let us now suppose that our state $\ket{\{\vb{x}\}}$ has $N_L$ atoms in the left arm and $N_R$ atoms in the right arm. Similarly let us take $\ket{\{\vb{x}'\}}$ to have $N_L'$ atoms in the left arm and $N_R'$ atoms in the right arm. 

Since \cref{lambda-d,lambda-u} are manifestly invariant under particle relabeling, let us order the particle labels such that $\vb{x}_1 \ldots \vb{x}_{N_L} = \vb{x}_L$ and $\vb{x}_{N_L+1} \ldots \vb{x}_N = \vb{x}_R$ (and similarly for the primed case).  It is convenient to align the kets in this form:
\begin{equation}
    \begin{split}
        \ket{\{\vb{x}\}} =&\ket{L, L,\ldots,\ldots, L, L}\overbrace{\ket{R,\ldots, R}}^{N_R}\\
        \ket{\{\vb{x}'\}}=&\underbrace{\ket{L,\ldots, L}}_{N_L'}\ket{R,R ,\ldots,\ldots, R, R}~.
    \end{split}
\end{equation}
We can now ask what the possible outcomes of the terms in the sum above above. It is clear that every exponential can evaluate to one of three numbers:
\begin{enumerate}
    \item $1$.
    \item $\e^{+\iu \vb{q} \cdot \Delta \vb{x}}$. 
    \item $\e^{-\iu \vb{q} \cdot \Delta \vb{x}}$.  
\end{enumerate}
We define the difference in path length via $\Delta \vb{x} = \vb{x}_L- \vb{x}_R$. $\mathbf{q}$ is the momentum transferred by the probe. Therefore, the decoherence kernel [i.e., term between squared brackets in \cref{lambda-d}] is given by
\begin{equation}
    \begin{split}
    &\frac12\bigg(\sum_{ij} 2\e^{\iu \vb{q}\cdot(\vb{x}_i -\vb{x}_j')} -\e^{\iu \vb{q}\cdot(\vb{x}_i -\vb{x}_j)} - \e^{\iu \vb{q}\cdot(\vb{x}'_i -\vb{x}'_j)} \bigg) \\ & 
    = \frac12\bigg(A_1  + A_+ \e^{+\iu \vb{q} \cdot \Delta \vb{x}} + A_- \e^{-\iu \vb{q} \cdot \Delta \vb{x}}\bigg)~.
    \end{split}
\end{equation}
Evaluating the coefficients $A_1$, $A_+$, and $A_-$ is a straightforward combinatorics problem.
We may without loss of generality, focus on the case where $n\geq 0$,
\begin{align}
    \begin{split}
    A_1 &= 2 N_L'N_L+2 N_R'N_R \\ & \quad -( N_L^2 + N_R^2 + N_L'^2 + N_R'^2)~, \\
        A_+ &= 2  N_L N_R' -( N_L N_R + N_L' N_R') ~, \\
       A_- &= 2  N_L' N_R - ( N_L N_R + N_L' N_R') ~.
   \end{split}  
\end{align}
As discussed in the main text, it is convenient to introduce the atom asymmetry
\begin{equation}
    n=N_L-N_L'=N_R -N_R'~,
\end{equation}
which characterizes how different the state $\ket{\{\vb{x}\}}$ is from the state $\ket{\{\vb{x}'\}}$, so that we may usefully express the previous expressions in terms of $n$,
\begin{equation}
    \begin{split}
    \label{expl-sum-A}
    A_1 &= -2n^2~,\\
    A_+ &= n(n-N) ~,\\
    A_- &= n(n+N) ~. 
    \end{split}
\end{equation}
It is important to note that all of these corrections vanish as $n\rightarrow 0$. Adding everything together we get the following expression for the decoherence kernel, 
\begin{equation}\label{eq:app_kernel_D}
    K_D = n^2 \qty[1-\cos(\vb{q}\cdot \Delta \vb{x})] - \iu n N \sin(\vb{q} \cdot \Delta \vb{x})~.
\end{equation}

The unitary kernel [i.e., term between squared brackets in \cref{lambda-u}]
contains only a subset of the terms above. We find for this case, 
\begin{equation}
    \begin{split}
    &\frac{\iu}{2}\bigg(\sum_{ij} \e^{\iu \vb{q'}\cdot(\vb{x}_i -\vb{x}_j)} - \e^{\iu \vb{q'}\cdot(\vb{x}'_i -\vb{x}'_j)} \bigg) \\ & = \frac{\iu}{2}\bigg(B_1  + B_+ \e^{+\iu \vb{q'} \cdot \Delta \vb{x}} + B_- \e^{-\iu \vb{q'} \cdot \Delta \vb{x}}\bigg)~.
    \end{split}
\end{equation}
These constants can be taken from the relevant terms above, 
\begin{equation}
    \begin{split}
    B_1 &=  N_L^2 + N_R^2 - N_L'^2 - N_R'^2~,\\
    B_+ &=  N_L N_R - N_L'N_R' ~,\\
    B_- &=  N_L N_R - N_L'N_R' ~, 
    \end{split}
\end{equation}
which can be expressed in terms of $n$ as 
\begin{equation}
    \begin{split}
    \label{expl-sum-B}
    B_1 &= -2n(n+N-2N_L)~,\\
    B_+ &= n(n+N-2N_L) ~,\\
    B_- &= n(n+N-2N_L)~.
    \end{split}
\end{equation}
As for the decohering part, these corrections vanish as $n\rightarrow 0$. From these expression we find the following closed-form expression for the unitary kernel, 
\begin{equation}\label{eq:app_kernel_U}
    K_U = -\iu \, n (n + N - 2N_L) \qty[1-\cos(\vb{q'}\cdot \Delta \vb{x})]~. 
\end{equation}

From \cref{eq:app_kernel_D,eq:app_kernel_U}, one can show that the multi-particle density matrix remains Hermitian after scattering. Since the multi-particle density matrix prior to scattering, $\rho_A$, is Hermitian by definition, 
\begin{equation}
    \rho_A(\{\vb{x}\},\{\vb{x}'\}) = \rho_A(\{\vb{x}'\},\{\vb{x}\})^*   ~.
\end{equation}
Therefore, for $\rho_A(t)$ to remain Hermitian for $t>0$, we require that
\begin{equation}
    \lambda_{U,D}(\{\vb{x}\},\{\vb{x}'\}) = \lambda_{U,D}(\{\vb{x}'\},\{\vb{x}\})^*~.
\end{equation}
For consistency, when mapping $n\rightarrow-n$ one must hold $N_L + N_L'$ fixed. Under this map we have 
\begin{align}
    \rho_A(\{\vb{x}\},\{\vb{x}'\})  &\rightarrow \rho_A(\{\vb{x}'\},\{\vb{x}\})~,\\
    K_{U,D}(\{\vb{x}\},\{\vb{x}'\}) &\rightarrow K_{U,D}(\{\vb{x}'\}, \{\vb{x}\}) \\
    &\hspace{12pt}= K_{U,D}(\{\vb{x}\},\{\vb{x}'\})^*~.\nonumber
\end{align}
Since $\lambda_{U,D}(\{\vb{x}\},\{\vb{x}'\}) \propto K_{U,D}(\{\vb{x}\},\{\vb{x}'\})$ and all other factors in $\lambda_{U,D}$ are real. It follows that the multi-particle density matrix after scattering is manifestly Hermitian. 

\section{Reduced density matrices \label{app:reduced-rho}}
In the main body of the text we have made use of reduced density matrices. For an $N$-body system, the $M$-body reduced density matrix is defined as $\rho_M= {\rm Tr}_{N-M}[ \rho]$. In the case of an $N$-atom system, ${\rm Tr}_{N-M}$ is the trace over $N-M$ single-atom Hilbert spaces.

Let us now consider the matrix elements of a reduced density matrix. Without loss of generality, take $\mel{\mathcal{N}_L'}{\rho_\alpha}{\mathcal{N}_L}$, where the states $\bra{\mathcal{N}_L'}$ and $\ket{\mathcal{N}_L}$ have $\mathcal{N}_L'$ and $\mathcal{N}_L$ particles in the left arm, and $\alpha-\mathcal{N}_L'$ and $\alpha-\mathcal{N}_L$ particles in the right arm, respectively. The partial trace which defines $\rho_\alpha$ involves a sum over states with $N_L = \mathcal{N}_L + M_L$ and $N_L' = \mathcal{N}_L' + M_L$ where $M_L$ accounts for the number of particles in the left arm from the states being traced over. Using the definitions of the kernels [\textit{cf.}~\cref{eq:app_kernel_U,eq:app_kernel_D}], and the fact that the unitary kernel is linear in $N_L$, we find
\begin{equation}
\begin{aligned}\label{app:eq:short-hand}
      \mel{\mathcal{N}_L'}{\rho_\alpha}{\mathcal{N}_L}  & =\mel{\mathcal{N}_L'}{\left[ \ \sum_{M_L=0}^{N-\alpha} \binom{N-\alpha}{M_L} \mel{M_L}{\rho}{M_L} \right]}{\mathcal{N}_L} \\
      &= \frac{1}{2^N}\e^{f(n)} \e^{\iu 2n \mathcal{N}_L \tau} \sum_{M_L=0}^{N-\alpha} {N-\alpha \choose M_L} \e^{\iu 2n M_L \tau} \\
      & = \frac{1}{2^\alpha} \e^{f(n)} \qty(\frac{1+\e^{\iu 2n \tau}}{2})^{N-\alpha} \e^{\iu 2 n \mathcal{N}_L \tau} \, ,
\end{aligned}
\end{equation}
for a function $f(n)$ given by, 
\begin{equation}\label{app:eq:f}
        f(n) = -n^2 s - \iu n N \gamma - \iu n(N+n) \tau \, ,
\end{equation}
where $s$, $\gamma$ and $\tau$ are defined in \cref{eq:s,eq:gamma,eq:tau}.
Importantly, \cref{app:eq:short-hand,app:eq:f} provide compact expressions for reduced density matrix elements.

For sufficiently weak coupling between the probe and the atoms (or equivalently at early times), the impact of the novel phase shift $\tau$ can be inferred from \cref{app:eq:short-hand} for a particular observable. For example, consider the expectation value of the observable $\mathcal{O}_+$ [\textit{cf}.~\cref{eq:expectation_O+,eq:vis_phase,rho1-expl}]. Since this observable is a one-body observable, it depends on a one-body reduced density matrix [i.e., $\alpha = 1$ in \cref{app:eq:short-hand}]. The expansion for $\langle \mathcal{O}_+ \rangle$ is given by 
\begin{equation}
    \begin{aligned}
    \langle \mathcal{O}_+ \rangle &= \frac{N}{2} \Bigg \{ \qty(1+ \cos\phi)  -\qty(s \cos\phi + N \gamma\sin \phi) \\ & + \bigg[\frac{1}{2}\cos\phi   \big( s^2+N^2\gamma^2-(N-1)\tau ^2  \big)+ Ns \gamma \sin\phi \bigg] \Bigg \} \\
    & + \ldots ~,
    \end{aligned}
\end{equation}
where the first bracketed expression is $O(t^0)$, the second is $O(t^1)$ and the remaining pieces are $O(t^2)$. We see that $\tau$ neither imprints a phase, nor affects decoherence at $O(t^1)$, but influences observables at $O(t^2)$.

\section{Statistical fluctuations of one-body operators}\label{app:stat_fluct}
In this appendix we provide completely general formulas for the statistical fluctuations in our toy model of a two-mode interferometer. We begin with a formulas for $\langle \mathcal{O}^\eta_{1B} \rangle$ with $\eta$ a positive integer, and then describe how to construct arbitrary statistical moments. 

Using the property that $\mathcal{O}_i^2=\mathcal{O}_i$ we find, 
\begin{equation}
    \mathcal{O}_{1B}^\eta = \sum_{\alpha = 1}^\eta C(\alpha) \sum_{\{i\}_{\rm dist.}} \mathcal{O}_{i_1}\mathcal{O}_{i_2} \ldots \mathcal{O}_{i_\alpha} ~.     
\end{equation}
The notation $\{i\}_{\rm dist.}$ refers to the set of distinct indices satisfying $i_1\neq i_2 \neq ... \neq i_\alpha$, and the product of operators should be understood as a tensor product. Since the expectation value $\langle \mathcal{O}_{i_1}\ldots \mathcal{O}_{i_\alpha} \rangle$ is invariant under permutations of the indices, we can replace the sum over the set $\{i\}_{\rm dist.}$ by $N(N-1)(N-2)\ldots (N-(\alpha-1))=N!/(N-\alpha)!$. We then find 
\begin{equation}
    \langle \mathcal{O}_{1B}^\eta \rangle = \sum_{\alpha = 1}^\eta C(\alpha) \frac{N!}{(N-\alpha)!} \langle \mathcal{O}_{i_1}\mathcal{O}_{i_2} \ldots \mathcal{O}_{i_\alpha} \rangle ~.
\end{equation}
The coefficients $C(\alpha)$ can be constructively determined using the Bookkeeper's Rule, however it is more efficient for $\eta \gtrsim 4$ to determine them iteratively using the sum rule $\sum_{\alpha} C(\alpha) N!/(N-\alpha)!=N^\eta$; for $3\leq \eta \leq 7$ the coefficients $C(\alpha)$ are shown in \cref{tab:C_alpha}. Next, when computing the expectation value we make use of the identity
\begin{equation}
    \langle \underbrace{\mathcal{O}_{i_1}\mathcal{O}_{i_2} \ldots \mathcal{O}_{i_\alpha}}_{\text{all indices distinct}}\rangle = {\rm Tr} \qty[ \mathcal{O}_{i_1}\mathcal{O}_{i_2} \ldots \mathcal{O}_{i_\alpha} \rho_\alpha  ] ~.
\end{equation}
This gives 
\begin{equation}\label{app:eq:exp_Oeta}
    \langle \mathcal{O}_{1B}^\eta \rangle =\sum_{\alpha = 1}^\eta C(\alpha) \frac{N!}{(N-\alpha)!} {\rm Tr}\qty[\mathcal{O}_{i_1}\mathcal{O}_{i_2} \ldots \mathcal{O}_{i_\alpha} \rho_\alpha]~.
\end{equation}
For $\mathcal{O}_{i} = \ket{A_i}\bra{A_i}$, where $\ket{A_i} = \frac{1}{\sqrt{2}}\qty( \ket{L_i} + \e^{\iu \phi} \ket{R_i})$, and using  the binomial distribution, the trace in \cref{app:eq:exp_Oeta} takes the form 
\begin{equation}
    \begin{aligned}
     {\rm Tr}\qty[\mathcal{O}_{i_1}\mathcal{O}_{i_2} \ldots \mathcal{O}_{i_\alpha} \rho_\alpha] = \frac{1}{2^\alpha}& \sum_{N_L,N_L'}  { \alpha\choose N_L} {\alpha\choose N_L'} \\ & \times \e^{\iu n \phi} \mel{N_L'}{\rho_\alpha}{N_L} \, ,
     \end{aligned}
\end{equation}
where we made use of the short-hand notation for the matrix elements of a reduced density matrix. 
It is convenient to re-express this as a sum over $n=N_L-N_L'$, 
\begin{equation}
    \begin{split}
    \sum_{N_L=0}^\alpha \sum_{N_L'=0}^\alpha = \sum_{n=-\alpha}^\alpha \sum_{N_L=0}^\alpha \sum_{N_L'=0}^\alpha \delta_{N_L-N_L',n} ~,
    \end{split}
\end{equation}
We may treat $n=0$, $n<0$, and $n>0$ separately. The results at $n<0$ are just the complex conjugate of $n>0$ and so we can further restrict to $n=0$ and $n>0$. For $n=0$, we simply set $N_L=N_L'$ and evaluate the sum. For $n>0$ the sum over $N_L$ runs from $N_L=n$ to $N_L=N$. By making use of \cref{app:eq:short-hand} one finds 
\begin{align}
     &{\rm Tr}\qty[\mathcal{O}_{i_1}\mathcal{O}_{i_2} \ldots \mathcal{O}_{i_\alpha} \rho_\alpha] =  \frac{1}{2^\alpha} \sum_{N_L=0}^\alpha { \alpha\choose N_L}^2 \\ & +  \frac{1}{2^\alpha}\sum_{n=1}^\alpha \sum_{N_L=n}^{\alpha} { \alpha\choose N_L} {\alpha\choose N_L'}\qty[ \e^{\iu n \phi} \mel{N_L'}{\rho_\alpha}{N_L} + {\rm c.c.}]~,\nonumber 
\end{align}
where $N_L' =N_L-n$. 

In general, it is possible to find analogous formulas for one-body observables of the type $\ket{A_i} =  a_L \ket{L_i} + a_R \ket{R_i}$, where $a_L, a_R \in \mathbb{C}$ and $|a_L|^2 + |a_R|^2 = 1$. In this case, these formulas  simply involve appropriate powers of $a_L$ and $a_R$ as would occur when expanding $(a_L+a_R)^{\alpha}$ using the binomial distribution.

\begin{table}[]
    \centering
    \begin{tabular}{cccccccc} \toprule
    {~$\eta$~}  & {~$C(1)$~} & {~$C(2)$~} & {~$C(3)$~} & {~$C(4)$~} & {~$C(5)$~} & {~$C(6)$~} & {~$C(7)$~}\\ \midrule
    3  & 1 & 3 & 1 & -- & -- & -- & -- \\
    4  & 1 & 7 & 6 & 1  & -- & -- & --  \\
    5  & 1 & 15 & 25 & 10 & 1 & -- & -- \\
    6  & 1 & 31 & 90 & 65 & 15 & 1 & --\\
    7  & 1 & 63 & 301 & 350 & 140 & 21 & 1  \\
    \bottomrule
    \end{tabular}
    \caption{Coefficients $C(\alpha)$ necessary to compute $\langle \mathcal{O}_{\rm 1B}^\eta \rangle $ for $\eta\in\{3,4,5,6,7\}$.}
    \label{tab:C_alpha}
\end{table}
\end{appendix}




\bibliography{biblio.bib}

\end{document}